\newcommand{\newmath}[2]{\newcommand{#1}{\mbox{${#2}$}}}
\newcommand{\be}{\begin{equation}}
\newcommand{\ee}{\end{equation}}
\newcommand{\bea}{\begin{eqnarray}}
\newcommand{\eea}{\end{eqnarray}}

\newcommand{\refeq}[1]{equation (\ref{eq:#1})}

\newcommand{\m}{\mu}

\newcommand{\whichd}{ { d } }
\newcommand{\cG}{ { \mathcal{G} } }
\newcommand{\cF}{ { \mathcal{F} } }

\newcommand{\half}{ { \frac{1}{2} } }

\newcommand{\dtotal}[2]     { { \frac{\whichd  #1}{\whichd  #2} } }
\newcommand{\dpartial}[2]   { { \frac{\partial #1}{\partial #2} } }
\newcommand{\dat}[3]        { { \left. \dtotal{#1}{#2}\right|_{#3} } }
\newcommand{\datpartial}[3] { { \left. \dpartial{#1}{#2}\right|_{#3} } }
\newcommand{\ddat}[1]       { { \dat{#1}{\rho}{\rho_d} } }
\newcommand{\ddatpartial}[1] { { \datpartial{#1}{\rho}{\rho_d} } }
\newcommand{\ddatpartialm}[1] { { \datpartial{#1}{\rho}{m_d, \rho_d} } }
\newcommand{\dmatpartial}[1] { { \datpartial{#1}{m}{m_d, \rho_d} } }
\newcommand{\ddGone}{ { \ddat{\cG_{1}} } }
\newcommand{\ddGtwo}{ { \ddat{\cG_{2}} } }
\newcommand{\ddGthree}{ { \ddat{\cG_{3}} } }
\newcommand{\dGAA}{ { \frac{\whichd \cG_{AA}}{\whichd \rho} } }
\newcommand{\dGBB}{ { \frac{\whichd \cG_{BB}}{\whichd \rho} } }
\newcommand{\dGAB}{ { \frac{\whichd \cG_{AB}}{\whichd \rho} } }

\newcommand{\dGs}{  { \ddatpartial{\cG^{*}}  } }

\newmath{\Fabij}{ { F_{\alpha \beta \,i j} } }

\newcommand{\rone}{ { \vec{r}_{1} } }
\newcommand{\rtwo}{ { \vec{r}_{2} } }
\newcommand{\rplain}{ { \vec{r} } }
\newmath{\Ri}{ { \vec{R}_i } } 
\newmath{\Rj}{ { \vec{R}_j } } 
\newcommand{\binv}{ { \beta^{-1} } }
\newcommand{\suma}{\sum_{\alpha}}
\newcommand{\sumb}{\sum_{\beta}}
\newcommand{\sumab}{\sum_{\alpha \, \beta}}
\newcommand{\sumi}{ {\sum_{i}} }
\newcommand{\sumiN}{ { \sum_{i=1}^{N} } }
\newcommand{\sumij}{\sum_{i \, j}}
\newcommand{\intv}[1]{ {\int_{V} d #1 \,} }
\newcommand{\intui}[1]{ {\int_{U_i} d #1 \,} }
\newcommand{\rhoar}[1]{ {\rho_{\alpha}(#1)} }
\newcommand{\rhobr}[1]{ {\rho_{\beta}(#1)} }
\newmath{\rhoaz}{\rho_{\alpha \, 0}}
\newmath{\rhobz}{\rho_{\beta \, 0}}
\newmath{\Da}{D_{\alpha}}
\newcommand{\Na}{N_{\alpha}}
\newmath{\Nd}{ { N_d } }
\newmath{\lai}{\lambda_{\alpha \, i}}
\newmath{\lAi}{ { \lambda_{A \, i} } }
\newmath{\lBi}{ { \lambda_{B \, i} } }

\newmath{\lAd}{\lambda_{A \, d}}
\newmath{\lBd}{\lambda_{B \, d}}
\newmath{\gai}{\gamma_{\alpha \, i}}
\newmath{\gbj}{\gamma_{\beta \, j}}
\newmath{\gAi}{\gamma_{A \, i}}
\newmath{\gBi}{\gamma_{B \, i}}
\newmath{\gAj}{\gamma_{A \, j}}
\newmath{\gBj}{\gamma_{B \, j}}
\newcommand{\gad}{\gamma_{\alpha \, d}}

\newmath{\gAd}{\gamma_{A \, d}}
\newmath{\gBd}{\gamma_{B \, d}}
\newmath{\gA}{\gamma_{A}}
\newmath{\gB}{\gamma_{B}}

\newcommand{\cai}{c_{\alpha \, i}}
\newcommand{\cbj}{c_{\beta \, j}}

\newcommand{\cb}{c_{\beta}}
\newcommand{\cAi}{c_{A \, i}}
\newcommand{\cBi}{c_{B \, i}}
\newcommand{\cAd}{c_{A \, d}}
\newcommand{\cBd}{c_{B \, d}}

\newmath{\Cabrr}{ C_{\alpha \beta}^{(2)} (\rone,\rtwo) }
\newmath{\Cabzr}{ C_{\alpha \beta}^{(2)} (\vec{0},\rtwo) }
\newcommand{\CabrrTerm}{\half \sumab \intv{\rone} d \rtwo \,
  \Cabrr \rhoar{\rone} \rhobr{\rtwo} }

\newcommand{\Cabijform}[3]{ { C_{#1 #2 \, i j}^{(2)} (#3) } }
\newcommand{\Cabij}{ {C_{\alpha \beta \, i j}^{(2)} (\rho)} }
\newcommand{\sumkcform}[1]{ {
  \sum_{k} \frac{\partial^{2} #1 }{\partial x_{k}^{2}} } }
\newcommand{\sumkc}{ { \sumkcform{\Cabij} } }

\newmath{\Cij}{ {C_{i j}^{(2)} (\rho)} }
\newmath{\Cijv}{ {C_{i j}^{(2)} (v)} }
\newcommand{\CijTerm}{ { \half \sumiN \sum_j m_i m_j \Cij } }

\newmath{\halfPmd}{ { \left( \half + m_d \right) } }   
\newmath{\halfMmd}{ { \left( \half - m_d \right) } }   
 
\newcommand{\lnlpart}[2]{ {
  \ln \left( \frac{\half + #1}{\left( \pi / #2 \right)^{3/2}} \right) 
                        } }
\newcommand{\lnrpart}[2]{ {
  \ln \left( \frac{\half - #1}{\left( \pi / #2 \right)^{3/2}} \right) 
                        } }
\newcommand{\lnwhole}[4]{ { 
  #1 \left\{ \left( \half + #2 \right) \lnlpart{#2}{#3} + \left(
\half - #2 \right) \, \lnrpart{#2}{#4} \right\}
                        } }

\newcommand{\lnterm}{ \lnwhole{\sumiN}{m_i}{\gAi}{\gBi} }

\newcommand{\dglnplus}[3]{ {
   \left( \half + #1 \right) \ln \left( \frac{#2}{#3} \right)     } }
\newcommand{\dglnminus}[3]{ {
   \left( \half - #1 \right) \ln \left( \frac{#2}{#3} \right)     } }

\newcommand{\mlnplus}{ {  \left( \half + m_i \right) 
   \ln \left( \frac{\half + m_i}{\half + m_d} \right)     } }
\newcommand{\mlnminus}{ {  \left( \half - m_i \right) 
   \ln \left( \frac{\half - m_i}{\half - m_d} \right)     } }
\newcommand{\mlnterm}{ { 
  \sumiN \left\{ \mlnplus + \mlnminus \right\}
                    } }
\newcommand{\glnplus}{ {  \left( \half + m_i \right) 
   \ln \left( \frac{\gAi}{\gAd} \right)     } }
\newcommand{\glnminus}{ {  \left( \half - m_i \right) 
   \ln \left( \frac{\gBi}{\gBd} \right)     } }
\newcommand{\glnterm}{ { 
  \frac{3}{2} \sumiN \left\{ \glnplus + \glnminus \right\}
                    } }

\documentstyle[aps]{revtex}
\begin{document}
\title{Volume changes in binary alloy ordering,\\
a binary classical density functional theory approach}
\author{David L. Olmsted}
\address{
The Martin Fisher School of Physics,\\ Brandeis University\\
Waltham, MA 02254, USA}
\date{June, 1998}
\maketitle
\begin{quotation}
\noindent{The chemical ordering transition in a binary alloy is
examined using classical density functional theory for
a binary mixture. The ordered lattice is assumed to be
obtained from the disordered lattice by a volume
change only, as in L1$_2$ ordering from an face centered cubic
chemically disordered crystal. Using the simplest possible
approach, second order truncation of the expansion, 
non-overlapping Gaussian distributions at the sites, and
expansion of the correlation functions about the sites, a
very tractable expansion is obtained. Under these assumptions
the expansion consists of the same terms as the lattice
gas formalism where the lattice is implicitly taken as fixed,
plus additional interaction terms, and an additional
entropy term. This additional entropy term represents a 
lowest order approximation to the vibrational entropy change.}
\end{quotation}
\section{Introduction.}
Chemical ordering transitions in alloys have been studied by
a variety of methods, including a lattice gas analogue of
classical density functional theory
\cite{gyorffy83,nieswand,seok97,ni3v}.
Since lattice changes also occur on ordering it would
be valuable to be able to include them along with the chemical
changes in a similar approach. One approach would involve a set
of variables including a global elastic strain tensor along with
chemical occupation variables\cite{elastic}. A question
that arises in such an approach is the appropriate form for
the ideal (non-interacting) free energy. It has been shown that
the occupation variables alone can be treated as a complete system, 
and that the ideal free energy is the ideal mixing entropy.
The occupation variables, plus a global strain tensor, are not
a complete system, however. One consistent set of variables
is the position and momentum variables for a binary mixture.
The approach taken here to developing a lattice gas plus
strain tensor formulation of ordering is to look to the
classical density functional theory of the binary mixture,
and simplify it to the lattice. Since the part of the strain tensor
that most directly affects the entropy is the volume, the other
terms will be ignored here. 

Several forms of classical density functional theory have been
used to study the freezing of binary liquids. Since the concern
here is the ``entropy'' terms, the simplest form of the
theory will be used. Here instead of looking at the freezing of
a liquid, the free energies of two solid structures (the disordered, 
and the (partially) ordered) are compared. A similar approach has
been used by Sengupta, Krishnamurthy and Ramakrishnan
to study the fcc-bcc interface\cite{sengupta94}, and a form of
density functional theory has been used to study the ordering of 
hard sphere mixtures\cite{dentonUPB}. 
\section{Classical DFT for multiple species.}
Classical density functional theory of mixtures has been used
to study the freezing of binary hard sphere and Lennard-Jones
fluids\cite{barrat86,barrat87,smithline87,rick89,zeng90,denton90}.

Consider a system with a fixed volume $V$, and $m$ species of classical
particles. [The primary interest herein is $m \! = \! 2$.] 
The partition function
is computed in the grand canonical ensemble at fixed values of $V$, the
temperature, $T$,
and the chemical potentials, \mbox{$\m_{\alpha},\, \alpha = 1,\ldots,m$}.

Classical DFT guarantees the existence of a functional $\Omega$ of the
average densities $\rhoar{\rplain}$ which is minimized by the
equilibrium average densities, and which evaluated at the equilibrium
densities is equal to the grand potential. For a non-interacting
system this can be computed explicitly and is\cite{barrat87}:
\be
\Omega_{ideal} [ \{ \rho_{\alpha} \} ] = 
    \binv \suma \intv{\rplain} \rhoar{\rplain}
    \left( \ln ( \lambda_{\alpha}^{3} \rhoar{\rplain} ) 
           - \beta \m_{\alpha} - 1\right), \label{eq:ideal}
\ee
where $\beta = 1/(k_B T)$, and $\lambda_{\alpha}$ is the de Broglie
thermal wavelength of species $\alpha$. 
Note that \refeq{ideal}
is simply the sum over the species of the ideal free
energy of each species.

Let $\beta^{-1} V \Phi [ \{ \rho_{\alpha} \} ] \equiv 
 \Omega [ \{ \rho_{\alpha} \} ]
- \Omega_{ideal} [ \{ \rho_{\alpha} \} ]$ and expand $\Phi$ about
some particular uniform (liquid)
state, with densities $\{\rhoaz\}$.
Truncating at second order in the difference in densities, 
$\delta\rho_{\alpha}$ 
(see \cite{barrat86,barrat87,smithline87,rick89,zeng90,denton90} for more
careful alternatives), and letting $ N_{\alpha} \equiv \intv{\rplain} \,
\rhoar{\rplain} $ one obtains as the expansion of the mixed functional
for the grand potential:
\bea
\beta \Omega[\{\rho_{\alpha} \}] & = & V \Phi[\{\rhoaz\}]  
                                             \nonumber \\ & & \mbox{}     
   +  \suma \intv{\rplain} \rhoar{\rplain} 
 \left( \ln \left( \lambda_{\alpha}^3 \rhoar{\rplain} \right) - 1 \right) 
                                             \nonumber \\ & & \mbox{}
   - \suma \beta \, \mu_{\alpha}  \Na           \nonumber \\ & & \mbox{}
   + \suma \intv{\rplain} \, C_{\alpha}^{(1)} 
        \left( \rhoar{\rplain} - \rhoaz \right)
                                             \nonumber \\ & & \mbox{}
   + \half \sumab \intv{\rone} d \rtwo \, \Cabrr 
     \left( \rhoar{\rone} - \rhoaz \right) 
     \left( \rhobr{\rtwo} - \rhobz \right)  \nonumber \\ & & \mbox{}
   + \mbox{O($\delta\rho_{\alpha}^3$)}.
\eea
Multiplying out the products, and collecting constant and linear
terms, this can be rewritten as:
{\samepage
\bea
\beta \Omega & \approx & 
           \suma \intv{\rplain} \rhoar{\rplain} \ln \rhoar{\rplain} 
                                           \nonumber \\ & & \mbox{}
         + V f_0                           \nonumber \\ & & \mbox{}
         + \suma \Da \Na                   \nonumber \\ & & \mbox{}
         + \CabrrTerm,              \label{eq:omega_rho}               
\eea
where
\bea
  f_0 & = & \Phi[\{\rhoaz\}]
          - \suma C_{\alpha}^{(1)} \rhoaz
          + \half \sumab \rhoaz \rhobz \intv{\rtwo} \Cabzr \\
  \Da & = & - \beta \, \mu_{\alpha} + \ln ( \lambda_{\alpha}^{3} ) 
            - 1
            + C_{\alpha}^{(1)}
            - \sumb \rhobz \intv{\rtwo} \Cabzr
\eea
}
By assuming that the uniform fluid at a particular total density and
set of concentrations is a local minimum of $\Omega$, \Da\ could be
eliminated in
favor of the liquid densities $\{ \rhoaz \}$. However, here the
task is to compare trial ordered solids with the disordered solid at
temperatures well below the melting point.
The \Da\ will therefore by retained for now, and later eliminated 
using the disordered solid as the reference state.

The constant $f_0$ drops out of the difference in grand potential
between two states, and so is irrelevant to determining the
transition state and temperature. It can be determined in terms
of the pressure of the reference disordered state.

\Cabrr\ is symmetric in $\alpha \, \beta$,
and depends only on $\rtwo - \rone$. (Note that for fixed 
$\left\{ \m_{\alpha} \right\}$ these liquid direct-correlation 
functions are needed as a function of temperature in a range
including the ordering temperature. In many systems this
would involve extrapolation beyond reasonable temperatures for
the liquid, and so this computation would not be feasible as
written. The purpose here, however, puts more emphasis on
understanding the appropriate form of the expansion for the 
solid, than on a computation using actual liquid data. Given
the appropriate form of the expansion stated in terms of
the occupation variables and strain tensor, other sources
for the parameters are likely to be more suitable to actual
numerical work.)
\section{Density Ansatz for Disordered and Ordered Solid}
Consider a fixed volume $V$. A solid is described in
terms of a set of real-space lattice vectors $ \{ \Ri \} $, with
$N$ sites in the volume $V$. Making the simplifying assumption that
there is exactly one
atom per site, $N = \suma \Na$. In general $\Omega$ depends
on $\{ \Ri \}$, and an assumption needs to be made as to what
possible lattices are being considered. To keep the situation as
simple as possible, assume that both the disordered solid
and the trial states for the ordered solid have the same
lattice structure, so that the $\{ \Ri \}$ depend only on $N$.
(Up to an arbitrary choice of origin for the lattice.)
So either $N$, or the lattice constant, or
the average total density can be considered as one parameter, 
and there is one constraint on the vector $\{\Na\}$. 
Think of $\rho \equiv N/V$
as the parameter, but $N$ will be written for $\rho V$
whenever convenient.

As in prior work on the freezing of hard-sphere
mixtures\cite{barrat86,barrat87,smithline87,rick89,zeng90,denton90}
, assume that the density of each species at a site is given by an
isotropic Gaussian distribution centered at the site. (In their
work on the fcc-bcc interface, Sengupta, et al., used 
a more general form of `anisotropic Gaussian distributions' and found
that there was significant anisotropy in the BCC density, but very
little in the FCC density.)  
For simplicity assume
that this is cut off at the Wigner-Seitz cell about the lattice
point. (And very quickly the further assumption will be made that
the Gaussian distributions
are sharply peaked enough that the integral over the cell is equal to
the integral over all space. Thus the assumption is essentially that the
Gaussian distributions are sharply enough peaked that there is no
overlap between the distributions at different sites.) 
Referring to the cell at
$\Ri$ as $U_{i}$ then in $U_{i}$
\be
  \rhoar{\rplain} = \lai \exp 
        \left\{ - \gai |\vec{r}-\vec{R}_{i}|^{2} \right\}
\ee
Here, for a binary alloy, $N$ and \{\lAi\}, \{\gAi\}, and \{\gBi\}\
($i=1..N$) parameterize
the density ansatz, while \{\lBi\}\ is then fixed by the assumption
of one atom per cell.
For the disordered solid, \lai\ and \gai\ are independent of $i$,
so the parameters are \Nd, \lAd, \gAd\ and \gBd. 

Define 
\be
  \cai \equiv \intui{\rplain} \lai 
       \exp \left\{ - \gai |\vec{r}-\Ri|^{2} \right\}.
\ee

At this point the limit $\gai \rightarrow 0, 
\lai \rightarrow { \rhoaz }$ gives the uniform binary
liquid. However, the states to be consider are those with
\gai\ large enough that $\int_{U_i}$ can be approximated by~$\int$.
For large enough \gai,
\be
  \cai \approx \lai \left( \frac{\pi}{\gai} \right) ^{3/2}.
\ee
\section{Results and Discussion}
The computation of $\Omega$ from equation~(\ref{eq:omega_rho}) for
the trial states described by the density ansatz proceeds as follows:
Inserting the density ansatz in the first term of (\ref{eq:omega_rho})
gives:
\be
  \suma \sumi \left\{ \cai \left( 
    \ln \left( \frac{\cai}{\left( \pi/\gai \right) ^{3/2}} \right)
    - 3/2 \right) \right\}
\ee
The last term depends not only on \Cabrr\ at the lattice points, but
also near the lattice points. Assuming that \Cabrr\ changes slowly
enough over the length scale $(1/\gai)^{1/2}$ that it can be 
expanded to second order, the result is:
\be
  \half \sumab \sumij \cai \cbj \left\{ 
     \Cabij 
 + \frac{1}{4} \left( \frac{1}{\gai} + \frac{1}{\gbj} \right)
   \sumkc
                                           \right\},
\ee
where
\be
  \Cabij \equiv C_{\alpha \beta}^{(2)} (\Ri(\rho),\Rj(\rho)).
\ee
{\samepage
Thus:
\nopagebreak
\bea
\beta \Omega & \approx & 
  \suma \sumi \left\{ \cai
    \ln \left( \frac{\cai}{\left( \pi/\gai \right) ^{3/2}} \right) \right\}
                                           \nonumber \\ & & \mbox{}
  + V f_0                                  \nonumber \\ & & \mbox{}
  + \suma \left( \Da - \frac{3}{2} \right) \Na
                                           \nonumber \\ & & \mbox{}
  + \half \sumab \sumij \cai \cbj \Cabij   \nonumber \\ & & \mbox{}
  + \half \sumab \sumij \cai \cbj \left( \frac{1}{4} \right) 
      \left( \frac{1}{\gai} + \frac{1}{\gbj} \right)
      \Fabij(\rho),
              \label{eq:omega_ab}               
\eea
where
\bea
  \Fabij & \equiv & \sumkc.
\eea
}
The equilibrium value of $\gai$ in this approximation is:
\be
  \gai = \frac{1}{6} \sum_{\beta} \sum_j \cbj \Fabij(\rho).
                                    \label{eq:gamma_i}
\ee
Upon substituting (\ref{eq:gamma_i}) into (\ref{eq:omega_ab}),
the final term cancels with the 3/2 in the third term. This
substantially simplifies the algebra, so minimizing
with respect to the \gai\ this is:
{\samepage
\bea
\beta \Omega & \approx & 
  \suma \sumi \left\{ \cai
    \ln \left( \frac{\cai}{\left( \pi/\gai \right) ^{3/2}} \right) \right\}
                                              \nonumber \\ & & \mbox{}
    + V f_0                                   \nonumber \\ & & \mbox{}
    + \suma \Da \Na                   
                                              \nonumber \\ & & \mbox{}
    +  \half \sumab \sumij \cai \cbj \Cabij 
                                              \nonumber \\
  \gai & = & \frac{1}{6} \sum_{\beta} \sum_j \cbj \Fabij(\rho).
                               \label{eq:omega_ab_d_2}               
\eea
}
For a binary alloy $c_{B}$ can be eliminated. It will prove
helpful to emphasize the relationship between eliminating
$c_{A}$ or $c_{B}$ by using the more symmetrical 
$m \equiv c_{A} - \half$. Thus
$m$ ranges from $-\half$ to $\half$. Then, for a binary alloy,
\bea
\beta \Omega & \approx & 
    \lnterm                              \nonumber \\ & & \mbox{}
  + V f_0                                \nonumber \\ & & \mbox{}
  + N \left( D_2 + \cG_2(\rho) \right) 
  + N m_{ave} \left( D_1 + \cG_1(\rho) \right)    \nonumber \\ & & \mbox{}
  + \CijTerm
                                              \nonumber \\
  \gai & = & \frac{1}{6} \sum_{\beta} \sum_j \cbj \Fabij(\rho),
                      \label{eq:binary}
\eea
where
\bea
  m_{ave} & \equiv & \frac{1}{N} \sumiN m_i                \\
  \Cij & \equiv & C_{A A \, i j}^{(2)} (\rho)  +  C_{B B \, i j}^{(2)} (\rho)
                  - 2 C_{A B \, i j}^{(2)} (\rho)                \\
  \cG_1 & \equiv & \half \sum_j C_{A A \, i j}^{(2)} (\rho)
                  - C_{B B \, i j}^{(2)} (\rho)                  \\
  \cG_2 & \equiv & \frac{1}{8} \sum_j 
      C_{A A \, i j}^{(2)} (\rho)  + C_{B B \, i j}^{(2)} (\rho)
      +   2 C_{A B \, i j}^{(2)} (\rho)      
                                \label{eq:def_G_2}            \\
  D_2 & \equiv & \half \left( D_A + D_B \right)              \\
  D_1 & \equiv & \left( D_A - D_B \right)                    \\
  \Fabij & \equiv & \sumkc,
\eea
and additionally let:
\bea
  \cG_3 & \equiv & \sum_j C_{i j}^{(2)} (\rho)
     = \sum_j C_{A A \, i j}^{(2)} (\rho)  
        +  C_{B B \, i j}^{(2)} (\rho)  
        -  2 C_{A B \, i j}^{(2)} (\rho)           \\
  \cF_{\alpha \beta} & \equiv & \sum_j F_{\alpha \beta \, i j} (\rho)  .
\eea

{\samepage
Now consider a disordered state, specified by a
concentration $m$ and an overall density $\rho$. At at given
temperature, this state will be in equilibrium for a  pair
of chemical potentials $\mu_A$ and $\mu_B$. (Now buried in $D_1$, $D_2$.) 
For this disordered state, the mixed functional for the 
grand potential per unit volume is:
\bea
\frac{\beta \Omega}{V} & \approx & 
    \lnwhole{\rho}{m}{\gAd}{\gBd}           \nonumber \\ & & \mbox{}
  + f_0                                   \nonumber \\ & & \mbox{}
  + \rho \left( D_2 + \cG_2(\rho) \right) 
  + \rho \, m \left( D_1 + \cG_1(\rho) \right)    \nonumber \\ & & \mbox{}
  + \rho \half m^2 \, \cG_3(\rho) 
                                              \nonumber \\
  \gad(m,\rho) & = & \frac{1}{6} \sum_{\beta} \cb \, \cF_{\alpha \beta}(\rho).
             \label{eq:omega_ij_d}
\eea
Which can be minimized with respect to $\rho$ and $m$ to obtain
the equilibrium density and concentration at a particular
$\mu_A$ and $\mu_B$, or vice-versa. 
}

{\samepage
Taking the disordered 
state ($m_d$, $\rho_d$) as the reference (equilibrium) state, and
minimizing equation (\ref{eq:omega_ij_d}) to eliminate the
chemical potentials gives:\nopagebreak
\bea
-D_1 & = & 
       \cG_1(\rho_d) + m_d \cG_3(\rho_d)       \nonumber \\ & & \mbox{}
     + \lnlpart{m_d}{\gAd} - \lnrpart{m_d}{\gBd}     
                                               \nonumber \\ & & \mbox{} 
     + \frac{3}{2} \halfPmd \dmatpartial{\ln \gAd} 
     + \frac{3}{2} \halfMmd \dmatpartial{\ln \gBd}  
                          \label{eq:D_1}
\eea
\pagebreak
\bea
-D_2 & = & 
       \cG_2(\rho_d) + \rho_d \ddGtwo          \nonumber \\ & & \mbox{}
     + m_d \rho_d \ddGone                      \nonumber \\ & & \mbox{}
     - \half m_d^2 \cG_3(\rho_d) + \half m_d^2 \rho_d \ddGthree
                                               \nonumber \\ & & \mbox{}
     + \half \lnlpart{m_d}{\gAd} + \half \lnrpart{m_d}{\gBd}     
                                               \nonumber \\ & & \mbox{}
     - \frac{3}{2} \halfPmd m_d \dmatpartial{\ln \gAd} 
     - \frac{3}{2} \halfMmd m_d \dmatpartial{\ln \gBd} 
                                               \nonumber \\ & & \mbox{}  
     + \frac{3}{2} \halfPmd \rho_d \ddatpartialm{\ln \gAd} 
     + \frac{3}{2} \halfMmd \rho_d \ddatpartialm{\ln \gBd}    
                         \label{eq:D_2}       \\
\left. \frac{\beta \Omega}{V} \right|_{m_d,\rho_d} & = & 
     f_0                                       \nonumber \\ & & \mbox{}
   - \rho_d^2 \ddGtwo                      
   - m_d \rho_d^2 \ddGone                      
   - \half m_d^2 \rho_d^2 \ddGthree             \nonumber \\ & & \mbox{}
   - \frac{3}{2} \halfPmd \rho_d^2 \ddatpartialm{\ln \gAd} 
   - \frac{3}{2} \halfMmd \rho_d^2 \ddatpartialm{\ln \gBd}     
                              \nonumber  \\
  \gad & = & \frac{1}{6} \sum_{\beta} \cb \, \cF_{\alpha \beta}(\rho_d),
                              \label{eq:reference}
\eea
Where the formula for $D_1$ has been used in computing $D_2$.
}

For studying the ordering transition this
is the desired result for the disordered state. However, 
it is interesting to take a detour and ask what this says about
the disordered state.

One question to ask is: ``How does the density vary with
concentration at constant pressure?''. The pressure (for an
equilibrium state) is given by $-p = \Omega / V$. Thus:
\bea
-\beta p & = & f_0                             \nonumber \\ & & \mbox{}
   - \rho_d^2 \ddGtwo                      
   - m_d \rho_d^2 \ddGone                      
   - \half m_d^2 \rho_d^2 \ddGthree            \nonumber \\ & & \mbox{}
   - \frac{3}{2} \halfPmd \rho_d^2 \ddatpartialm{\ln \gAd} 
   - \frac{3}{2} \halfMmd \rho_d^2 \ddatpartialm{\ln \gBd},          
\eea
and switching back from $m$ to $c$ for a moment, this is:
\bea
\beta p & = & - f_0                    \nonumber \\ & & \mbox{}
   + \half \rho^2 \left\{
        c_A^2 \dGAA + c_B^2 \dGBB + 2 c_A c_B \dGAB 
                  \right\}             \nonumber \\ & & \mbox{}
   + \frac{3}{2} c_A \, \rho^2 \dpartial{\ln \gAd}{\rho} 
   + \frac{3}{2} c_B \, \rho^2 \dpartial{\ln \gBd}{\rho}. 
                         \label{eq: pressure}         \\
\cG_{\alpha \beta} & \equiv & 
           \sum_j C_{\alpha \beta \, ij}^{(2)}(\rho).
\eea
For a given $p$, $m$, and $T$, this equation implicitly gives the
equilibrium density of the disordered solid, assuming it exists.

Some insight into this equation can be gained by imagining that the
$\cG_{\alpha \beta}$ are dominated by nearest-neighbor contributions.
One then expects that for a single species the equilibrium density is
give by a density near the minimum of $\cG_{A A}$. (That is, the
nearest-neighbor distance in the solid is approximately equal to
distance at which the liquid's direct correlation function has its
main peak.) If the two species are significantly different in size,
it plausible to assume that the terms involving $\cG_{\alpha \beta}$
in (\ref{eq: pressure}) are more important than those involving
$\gad$. Suppose the latter are small. At the pressure where
$\beta p + f_{0} = 0$, the density corresponding to this pressure
is given by:
\be
 0  =  c_A^2 \dGAA + c_B^2 \dGBB + 2 c_A c_B \dGAB.  
\ee
And further simplifying by assuming 
$C_{A B}^{(2)}(\rplain) =
 \left( C_{A A}^{(2)}(\rplain) + C_{B B}^{(2)}(\rplain) \right) / 2,$
this reduces to 
\be
 0  =  c_A \dGAA + (1 - c_A)  \dGBB.  
\ee

Considering nearest-neighbor contributions only, and assuming
that both $C_{A A}^{(2)}(\rplain)$
and $C_{B B}^{(2)}(\rplain)$ can be expanded as quadratics about
their minima over the relevant region:
\bea
\cG_{AA}(r_{nn}) & = & K_{A} + \half k_A \left( r_{nn} - r_{A} \right)^2
               \nonumber \\
\cG_{BB}(r_{nn}) & = & K_{B} + \half k_B \left( r_{nn} - r_{B} \right)^2
               \label{eq:rnnquad}
\eea
where $r_{nn}$ is the nearest neighbor distance corresponding to
a given density, then equation (\ref{eq:rnnquad}) gives:
\be
  r_{nn} = \frac{c_{A} k_{A}}{c_{A} k_{A} + (1-c_{A}) k_{B}} r_{A} 
          + \frac{(1-c_{A}) k_{B}}{c_{A} k_{A} + (1-c_{A}) k_{B}} r_{B}.
\ee
And if $k_A = k_B$, this makes $r_{nn}$ linear in concentration.

Returning to the analysis of the ordering transition, the
next step is to compute the difference in $\Omega$ between
an trial state and an equilibrium disordered state. 
At the transition, both the reference disordered state and the
(partially) ordered state at the transition will be global minima of
$\Omega$, and the difference in $\Omega$ will be zero. Hopefully
the approximation to $\Omega$ is good enough, and the range
of trial states provide by the density ansatz is generous enough,
that the solution provided will approximate the actual transition.
Only trial states with $\gai$ given by the last line of
equation (\ref{eq:binary}) need to be considered. 
The trial states are therefore considered as a function of
$\rho$ and $\{m_i\}$. $m_d$, $\rho_d$,
$\gAd$, and $\gBd$ refer to the equilibrium state for the given
chemical potentials and temperature. Combining (\ref{eq:binary})
and (\ref{eq:D_1}) through (\ref{eq:reference}):
\bea
  \Delta \beta \Omega & \approx & \mlnterm
                                               \nonumber \\ & & \mbox{}
  + \glnterm
                                          \nonumber \\ & & \mbox{}
  - \frac{3}{2} \left( N - N_d \right) 
    \left\{    \halfPmd \rho_d \ddatpartialm{\ln \gAd} 
             + \halfMmd \rho_d \ddatpartialm{\ln \gBd}   \right\}
                                               \nonumber \\ & & \mbox{}
  - \frac{3}{2} N \left( m_{ave} - m_d \right) 
    \left\{  \halfPmd \dmatpartial{\ln \gAd} 
           + \halfMmd \dmatpartial{\ln \gBd}  \right\} 
                                               \nonumber \\ & & \mbox{}
  + N \left( \cG_2(\rho) - \cG_2(\rho_d) \right) 
         - \left( N - N_d \right) \rho_d \ddGtwo
                                               \nonumber \\ & & \mbox{}
  + N m_{ave} \left( \cG_1(\rho) - \cG_1(\rho_d) \right) 
         - \left( N - N_d \right) m_d \rho_d \ddGone
                                               \nonumber \\ & & \mbox{}
  + \half \sumiN \sum_j \left( m_i - m_d \right)
        \left( m_j - m_d \right) \Cij          \nonumber \\ & & \mbox{}
  + N \left( m_{ave} - \half m_d \right) 
        m_d \left( \cG_3(\rho) - \cG_3(\rho_d) \right)
                                               \nonumber \\ & & \mbox{}
  - \half \left( N - N_d \right) m_d^2 \rho_d \ddGthree
                                               \nonumber \\
  \gai & = & \frac{1}{6} \sum_{\beta} \sum_j \cbj \Fabij(\rho).
                                               \nonumber \\
                     \label{eq:diff_ij}
\eea
The first term is the ideal mixing entropy for a pure
Ising-like model. It is the same term that appears in the lattice
gas formalism. 
More interesting is the second term (corrected by the next two term
which cancel its linear parts) which also derives from the ideal part
of the free energy. $(\gad/\pi)^{-3/2}$ has units of volume, and 
represents the approximate volume over which the atom is likely to
be. Writing this term as
\bea
  \Delta \beta \Omega & \approx & \ldots           
  - \sumi \left\{ \dglnplus{m_i}{\gAi^{-3/2}}{\gAd^{-3/2}} 
                      + \dglnminus{m_i}{\gBi^{-3/2}}{\gBd^{-3/2}} \right\} 
   + \ldots,        \nonumber \\
   & = & \ldots
  - \sumi \left\{ \cAi \ln \left( \frac{\gAi^{-3/2}}{\gAd^{-3/2}} \right)
      + \cBi \ln \left( \frac{\gBi^{-3/2}}{\gBd^{-3/2}} \right) \right\}
   + \ldots,
\eea
it can be thought of as representing the difference in an entropy
term based on the average volume available to the atom to wander in.
The assumption of Gaussian distributions approximates each atom as
an independent oscillator from this point of view. We can therefor
consider this expression as a lowest non-zero order approximation to
the vibrational entropy difference.
Note that this ``entropy'' term, which
does not appear in the lattice gas formalism, does not directly
depend on the density. Rather it depends on $\gamma$, which 
is the variable describing the probability distribution for
an atom about its site, in our very simple density ansatz.
This makes sense. In the lattice gas model there is just an
occupation variable at a site, with no idea of a fluctuation
of an atom position about a site. The non-overlapping Gaussian
distributions we have adopted correspond to attaching an atom to
each site with an independent harmonic spring. (Where the
spring constant depends on both the overall density, and the
state of chemical order.) In this picture it is not surprising that
the entropy can be represented in term of the mixing entropy plus
an entropy term based purely on the effective volume an atom
occupies about its site. Thus this term represents a lowest-order
estimate of the vibrational entropy change on ordering. Even in
this simple approximation, however, this term depends on both
the overall density and the state of chemical order, as seen below.

The remaining terms can be rewritten as
\nopagebreak
\bea
  \Delta \beta \Omega \approx \ldots          
  & + & \half \sumiN \sum_j \left( m_i - m_d \right)
        \left( m_j - m_d \right) \Cij          \nonumber \\ \mbox{}
  & + & N \left( m_{ave} - m_d \right) 
        \left( \cG^{**}(\rho) - \cG^{**}(\rho_d) \right)
                                               \nonumber \\ \mbox{}
  & + & \half N \left( \cG^{*}(\rho) - \cG^{*}(\rho_d) \right) 
         - \left( N - N_d \right) \rho_d \dGs,
\eea
where
\bea
 \cG^{*}(\rho) & \equiv & 
               2 \cG_2(\rho) + 2 m_d \cG_1(\rho) + m_d^2 \cG_3(\rho)
                                               \nonumber \\
            & = & \cAd^2 \, \cG_{A A}(\rho) 
                   + 2 \cAd \, \cBd \, \cG_{A B}(\rho)
                   + \cBd^2 \, \cG_{B B}(\rho)
                                               \nonumber \\
 \cG^{**}(\rho) & \equiv & \cG_1(\rho) + m_d \cG_3(\rho)
                                               \nonumber \\
             & = & \cAd \, \cG_{A A}(\rho) 
                    + \left( \cBd - \cAd \right) \cG_{A B}
                            - \cBd \, \cG_{B B}(\rho).
\eea  
The first term is the second order term that appears in the
lattice gas formulation, and the $C^{(2)}$ are available for
many systems from experiment or first principles calculations.
Note that it is most conveniently
written in terms of the direct correlation function at the 
{\bf trial} lattice constant. 
The last term is O($(\rho - \rho_d)^2$) (the O($\rho - \rho_d$) portions
of the two sub-terms cancel on expanding $\cG^{*}(\rho)$ about $\rho_d$) and
can be estimated from the bulk modulus of the reference state.
The middle term is O($(m_{ave} - m_d)(\rho - \rho_d)$), and
can be estimated from the concentration dependence of the
density of the disordered alloy.

Equation (\ref{eq:diff_ij}) is the primary result. 
It is expressed in terms of the
desired variables, $\rho$ and $\{m_i\}$. Except for the
``vibrational entropy'' terms the coefficients can be estimated
using available information. 
It is used to
predict the transition from the disordered to the ordered state in
the same way as the lattice gas analogue of classical
density functional theory is. Given $\{\rho_d, m_d\}$,
the disordered state always minimizes the free energy, by construction.
As the temperature is lowered, there may be another minimum of
the free energy. The highest temperature where there is an ordered
state which minimizes the free energy, and for which the free energy
difference is zero is the predicted transition. If the ordered state
at the transition differs from the disordered state by a finite amount
the transition is predicted to be first order. For a second order
transition the ordered state will be equal to the disordered state
at the transition, but can be determined by the fact that there
will be ordered states different from the disordered states
that minimize the free energy, and have free energies lower than
the disordered state, for any temperature strictly less than
the transition temperature.

One difficulty exists. The vibrational entropy term is still
expressed in terms of the liquid partial direct correlation functions.
These are unlikely to be available and suitable for many
systems. It would be even more appropriate to use the partial
correlation functions of the solid, but these are also unlikely to be
available.  

What is $\gamma$ like in this theory? From equation (\ref{eq:gamma_i})
we have:
\be
  \gAi = \frac{1}{6} \left\{ \sum_j c_j \sumkcform{\Cabijform{A}{A}{\rho}}
          + \sum_j (1-c_i) \sumkcform{\Cabijform{A}{B}{\rho}} \right\}.
\ee
This talks about the curvature of the liquid partial direct
correlation functions. Note that
\be \sumkcform{\Cabijform{\alpha}{\beta}{\rho}} =
  C^{(2)\prime\prime}_{\alpha\beta}(r_{ij}(\rho))
   +\frac{2}{r_{ij}(\rho)} C^{(2)\prime}_{\alpha\beta}(r_{ij}(\rho))
\ee
At constant density this approximation is linear in concentration for
a disordered state.
The density dependence will be complicated, however. Considering
nearest neighbors only, we expect to be near the minimum of
$\Cabij$, subject to the competition between three different terms,
and the contributions of the further neighbors. etc. 
Over a small range the curvature might have a simple form, but
it eventually must go to zero in each direction.  
Because of the sum over the neighbor shells, the $\gai$ depend
on the state of chemical order as well. 

In what limit does this reduce to the lattice gas expansion?
For large bulk modulus the density change will be small. 
However even for fixed density, the dependence of $\gai$ on the
ordering will appear. Thus to regain the lattice gas formula, 
an additional assumption that $\gai=\gad$ as well as $\rho=\rho_d$
must be made. 
\section{Conclusions}
The above computation demonstrates how a truncated binary density
functional expansion has a form as the terms appearing in the
equivalent lattice gas expansion, plus additional terms.
These additional terms are the expected ``interaction''
terms from the added density variable, and an additional
ideal gas entropy term involving $\ln(\gamma)$ 
This computation is easily extended, under the same assumptions,
to a global strain tensor variable in place of the density variable.
The results are the same, except that a particular tensorial form
for the ``interaction'' terms involving the traceless portion
of the strain tensor is chosen from the possible forms,
because the $C^{(2)}$ depend on distance only and have no angular part.
(This will not be the case if the expansion is extended to $C^{(3)}$.)

In order to use this formalism in situations where the liquid
correlation functions are unavailable some assumption will
need to be made in regard to the `$\ln(\gamma)$' terms.
One way to proceed is to ignore equation (\ref{eq:gamma_i}) and
make an ad-hoc assumption for the $\gamma$'s. In a ``purely
harmonic'' approximation we could assume $\gamma$ to be a constant,
independent of density, concentration, and order. In this
case it drops out of \refeq{diff_ij} entirely.
A slightly more ambitious assumption would be to assume that for
the ``volume'' of the distribution scales
as the volume per atom. In this case the density dependence
of the $\ln(\gamma)$ term is numerically very small compared
to the other terms, so that in numerical solutions it would
be negligible.

This application of this formalism, including a general 
global elastic strain 
tensor rather than just the density as considered here, to 
Nickel-rich Nickel-Vanadium alloys is in progress, 
under the assumption that the `$\ln(\gamma)$' terms
are small\cite{elastic}. Work is also planned to apply the formalism to
the simpler case of a volume change only in Cu$_3$Au, where
estimates of the change in $\ln(\gamma)$ can be made from
embedded atom simulations.
\section*{Acknowledgments}
F. Pinksi and B. Chakraborty proposed the question partially answered
here, suggested this approach, and provided helpful discussions. 
B. Chakraborty was also kind enough
to provide a detailed reading of the manuscript. 
Helpful discussions with H. Krishnamurthy
are also gratefully acknowledged. 
This work was supported
in part by the National Science Foundation under grant DMR-9520923.


\begin{thebibliography}{10}
\bibitem{gyorffy83}
B.~L. Gy{\"{o}}rffy and G.~M. Stocks, Phys. Rev. Lett. {\bf 50},  374  (1983).

\bibitem{nieswand}
M. Nieswand, W. Dietrich, and A. Majhofer, 
    Phys. Rev. E {\bf 47},  718  (1993);
  M. Nieswand, A. Majhofer, and W. Dietrich, 
    Phys. Rev. E {\bf 48},  2521 (1993).
  
\bibitem{seok97}
C. Seok and D.~W. Oxtoby, J. Phys.:Condens. Matter {\bf 9},  87  (1997).

\bibitem{ni3v}
D.~L. Olmsted, F.~J. Pinski, J.~B. Staunton, 
  D.~D. Johnson, and B. Chakraborty,
  Short-range ordering in the Nickel-Vanadium system, in preparation.

\bibitem{elastic}
D.~L. Olmsted, F.~J. Pinski, and B. Chakraborty, work in progress.

\bibitem{sengupta94}
S. Sengupta, H.~R. Krishnamurthy, and T.~V. Ramakrishnan, 
  Europhys. Lett. {\bf 27},  587  (1994).

\bibitem{dentonUPB}
A.~R. Denton and N.~W. Ashcroft, unpublished, cited in [13].

\bibitem{barrat86}
J.~L. Barrat, M. Baus, and J.~P. Hansen, Phys. Rev. Lett. {\bf 56},  1063
  (1986).

\bibitem{barrat87}
J.~L. Barrat, M. Baus, and J.~P. Hansen, J. Phys. C {\bf 20},  1413  (1987).

\bibitem{smithline87}
S.~J. Smithline and A.~D.~J. Haymet, J. Chem. Phys. {\bf 86},  6486  (1987),
  erratum: {\bf 88}, 4104, (1988).

\bibitem{rick89}
S.~W. Rick and A.~D.~J. Haymet, J. Chem. Phys. {\bf 90},  1188  (1989).

\bibitem{zeng90}
X.~C. Zeng and D.~W. Oxtoby, J. Chem. Phys. {\bf 93},  4357  (1990).

\bibitem{denton90}
A.~R. Denton and N.~W. Ashcroft, Phys. Rev. A {\bf 42},  7312  (1990).
\end{thebibliography}
\end{document}